\pdfoutput=1
\documentclass{article}
\usepackage{graphicx, indentfirst, hyperref}
\usepackage[a4paper, margin = 3cm]{geometry}
\usepackage[numbers, sort&compress]{natbib}
\makeatletter
\let\mailmark\@fnsymbol
\makeatother

\newcommand*{\cf}{\emph{cf.}}
\newcommand*{\eg}{\emph{eg.}}
\newcommand*{\etc}{\emph{etc}}
\newcommand*{\prog}[1]{\emph{#1}}
\newcommand*{\figref}[1]{Figure \ref{fig:#1}}
\newcommand*{\secref}[1]{Section \ref{sec:#1}}
\let\thxmark\textsuperscript
\begin{document}
\title{%
	Beyond the EPICS:\\
	comprehensive Python IOC development with \prog{QueueIOC}%
}
\author{%
	Peng-Cheng Li\thxmark{1,2,3}, Xiao-Xue Bi\thxmark{1},
	Ying-Ke Huang\thxmark{1}, Dian-Shuai Zhang\thxmark{1},\\
	Xiao-Bao Deng\thxmark{1}, Qun Zhang\thxmark{1,4}, Ge Lei\thxmark{1,3},
	Gang Li\thxmark{1}, Yu Liu\thxmark{1,\mailmark{1}}%
}
\date{}
\maketitle
\begingroup
\renewcommand{\thefootnote}{\fnsymbol{footnote}}
\footnotetext[1]{\ Correspondence e-mail: \texttt{liuyu91@ihep.ac.cn}.}
\endgroup
\footnotetext[1]{\ %
	Institute of High Energy Physics, Chinese Academy of Sciences,
	Beijing 100049, People's Republic of China.%
}
\footnotetext[2]{\ %
	National Synchrotron Radiation Laboratory,
	University of Science and Technology of China,
	Hefei, Anhui 230029, People's Republic of China.%
}
\footnotetext[3]{\ %
	University of Chinese Academy of Sciences,
	Beijing 100049, People's Republic of China.%
}
\footnotetext[4]{\ %
	North China University of Technology,
	Beijing 100144, People's Republic of China.%
}

\section*{Abstract}

\paragraph{Keywords:}
EPICS; Python IOC; submit/notify pattern;
\prog{s6-epics}; software architecture.

\paragraph{Background and Purpose:}
Architectural deficiencies in EPICS lead to inefficiency in the
development and application of EPICS IOCs.  An unintrusive solution
is replacing EPICS IOCs with more maintainable and flexible Python
IOCs, only reusing the CA protocol of EPICS.  While there are libraries
like \prog{caproto} and \prog{PCASPy} that help to create Python IOCs,
they still feel insufficient for more complex requirements.

\paragraph{Methods:}
Noticing \verb|caput|, \verb|caget| and \verb|camonitor| are just
specialised combinations of requests/\linebreak[1]replies and notifications
in client-server communication, by combining barebone \prog{caproto}
and event loops like those in server-like programs, the \prog{QueueIOC}
framework for Python IOCs is created, which has the potential to
systematically reduce the development and maintenance cost of IOCs.

\paragraph{Results:}
Examples based on \prog{QueueIOC} are first given for workalikes of \prog%
{StreamDevice} and \prog{asyn}; also given are examples for ``sequencer''
applications, like those based on \prog{seq}, include monochromators, motor
anti-bumping and motor multiplexing.  A \prog{QueueIOC}-based framework for
detector integration is presented in an accompanying paper.  Also reported
is a simple but expressive architecture for GUIs, as well as software
to use with the \verb|~/iocBoot| convention which addresses some
issues we find with a similar solution based on \prog{procServ}.

\section{Introduction}\label{sec:intro}

The Experimental Physics and Industrial Control System (EPICS) is a
basis for accelerator control and beamline control at HEPS, the High Energy
Photon Source \cite{chu2018, liu2022a}.  The rich selection of EPICS
modules available has greatly facilitated device-control tasks at HEPS.
However, during these tasks, it also became apparent that EPICS has
some deficiencies inherent in its architecture (\cf\ \figref{ioc-arch}b),
which lead to inefficiency in development and application.  Our first
complaints with EPICS are about its conception of ``record links''
\cite{kraimer2018}.  EPICS record links have different types (input
links, output links and forward links), subtypes (database links, Channel
Access links \etc) and attributes (\eg\ \verb|NPP|, \verb|PP|, \verb|CA|,
\verb|CP| and \verb|CPP| for the ``Process Passive'' attribute).  Events
(hardware interrupts, user writing, periodic ``scanning'') result in record
``processing'', where the selection and ordering of records to be processed
are determined by these links according to a large and complex rule set.
The rule set cannot be decomposed into separate rules about types, subtypes
and attributes, which leads to considerable difficulty in learning.  And even
with the already complex rule set, there are still requirements that cannot
be naturally done, and a prime example for this is the \prog{motor} module.
In addition to the types and attributes of links, the selection and ordering
of records to be processed are also affected by the records themselves,
depending on them being ``synchronous'' or ``asynchronous''; however,
\verb|motor| records are neither synchronous nor asynchronous.

\begin{figure}[htbp]\centering
\includegraphics[width = 0.8\textwidth]{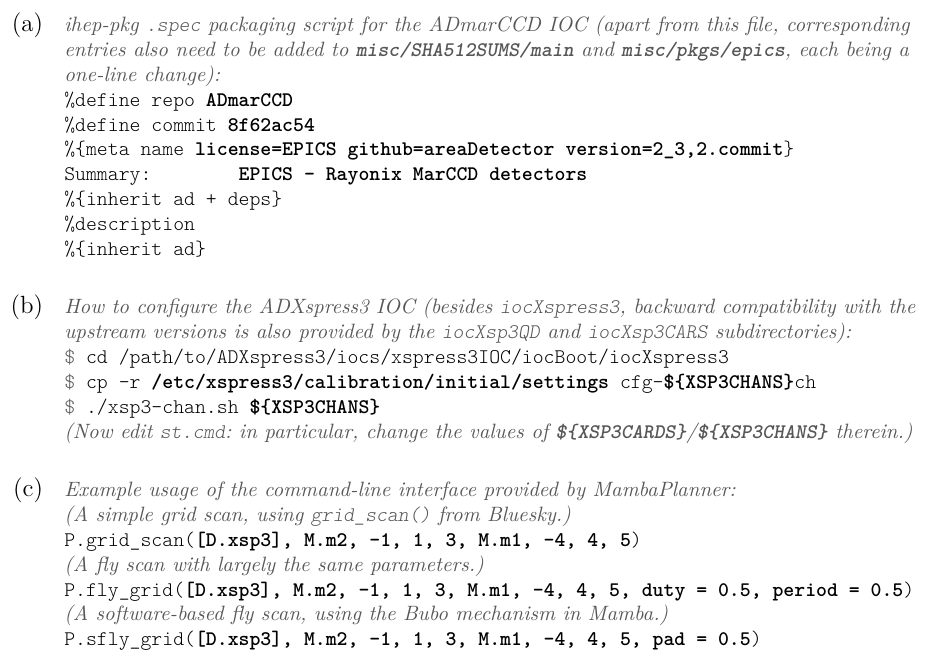}
\caption{Some code examples, with the essential information in bold.}
\label{fig:lower-bound}
\end{figure}

Apart from the link mechanism, EPICS records themselves are also quite
inexpressive: EPICS ``databases'', which are collections of records,
cannot be nested at runtime, which makes it very hard to abstract reusable
functionalities as pluggable sub-databases, even for simple requirements like
proportional-integral-derivative (PID) controllers.  One idiomatic alternative
is to implement customised records, \eg\ \verb|epid| from the \prog{std} module,
which involves a considerable amount of repeated code and quite an extent
of knowledge on EPICS internals.  Another common alternative is writing
``sequencers'' based on the \prog{seq} module, which is often error-prone
(\cf\ \secref{cases}).  The \verb|st.cmd| language in EPICS does not
have looping or conditional constructs, so we also cannot circumvent the
inexpressiveness of records on the \verb|st.cmd| layer.  A common practice
to work around this is to use external code generators, \eg\ \prog{iocbuilder}
\cite{abbott2011}, \texttt{makeDb.py} (from the \prog{ADGenICam} module) and
our own ones (\cf\ \prog{ADXspress3} in the next paragraph).  \verb|st.cmd| also
feels inconvenient to operators, further impeding user-oriented abstraction.
It is undoubted that none of these issues is fatal to EPICS: developers
can still produce usable, reliable EPICS modules, and users can still use
these modules to actually fulfill their needs; as has been summarised above,
the problems are about inexpressiveness and inconvenience.  Meanwhile, during
the construction of HEPS, we have found that improvements on expressiveness
and convenience can be of great value in boosting efficiency; here we
briefly give a few examples for this (\figref{lower-bound}).

The \prog{ihep-pkg} packaging framework \cite{liu2022a} now covers a larger
range of EPICS modules than that of the NSLS-II \prog{epicsdeb} repository
\cite{bnl2018}; meanwhile it is still easy to keep up to date, even in
comparison with recent works like \prog{installSynApps} \cite{derbenev2023}.
It provides support for CentOS 7, Rocky 8 and hopefully compatible
environments like RHEL 7/8; partial support is provided for Windows,
Debian/Ubuntu and Rocky/RHEL 9, mainly targeting scenarios where the C/C++
libraries (conventionally called ``\emph{SDKs}'' or software development kits)
from vendors mandate the use of these environments.  The \prog{ADXspress3}
module (\url{https://codeberg.org/CasperVector/ADXspress3})
merges functionalities from both of its upstream versions
(\url{https://github.com/quantumdetectors/xspress3-epics} and
\url{https://github.com/epics-modules/xspress3}), and is much easier
to maintain than both upstreams; meanwhile, to users it is also more convenient
to build and use.  The \prog{MambaPlanner} mechanism \cite{li2023} provides a
succinct, consistent command-line interface (CLI) for both step scans and fly
scans, encapsulating hairy details about hardware configuration, correctness
checks and data processing; while being convenient for debugging and advanced
usage, it also facilitates development of graphical user interfaces (GUIs)
oriented toward regular users.  As is shown in \figref{lower-bound}, they
have each approached some kind of \emph{complexity/cost lower-bound}, in the
sense that there is no obvious way to significantly simplify the code without
inordinately complicating other code modules or sacrificing code clarity.
In the examples above, the efficiency boosts in them can be in 1--2 orders
of magnitude, which also nicely coincide with the reduction of code needed to
do the same tasks.  Similarly, in other applications at HEPS, improvements
in succinctness also prove to coincide with improvements in efficiency,
even when the complexity lower-bounds have not yet been approached.  More
profoundly, as has been discussed recently \cite{li2025}, we believe that
the pursuit to minimise complexity/cost is a pursuit of the collaboration
and coevolution of artificial intelligence (AI) with human intelligence.

From the perspective above, we can give a summary of the problems
with EPICS: the essence of these problems is that its architecture
makes the complexities of EPICS modules vastly higher than the lower
bounds; the latter can often be estimated intuitively, just like in
\figref{lower-bound}.  We understand that the design of EPICS is deeply
affected by historical factors: the choice of record and links over a
full-fledged language, as well as the inexpressiveness of \verb|st.cmd|,
may be largely seen as results of hardware limitations on Versa Module Europa
(VME) and similar hardware platforms; record and links are also reminiscent
of the programming paradigm common with programmable logical controllers
(PLCs), PandABox \cite{christian2019} \etc.  Nevertheless, with the fast
growth of EPICS usage on PCs, it is also time to rethink about these design
decisions.  Of course, an option for completely new facilities is to embrace
ecosystems with less historical burdens, \eg\ \prog{Karabo} \cite{gories2023};
for facilities with more legacies, a more backward-compatible option is to
reuse the Channel Access (CA) protocol of EPICS, but internally use
more succinct tools.  Given the common programming practices in last decade
(especially those related to large scientific facilities), Python is an
obvious candidate for this task; in particular, we have chosen the pure-Python
\prog{caproto} library \cite{allan2021}, as it provides succinct interfaces for
both server-side and client-side programming with the CA protocol; alternatives
like \prog{PCASPy} (\url{https://github.com/paulscherrerinstitute/pcaspy})
may have issues like \url{https://github.com/pyepics/pyepics/issues/176}.
However, the officially expected usage of \prog{caproto} depends on Python's
\verb|async|/\verb|await| programming paradigm, which leads to additional
costs in learning and development, although it is very helpful in highly
concurrent application scenarios; in this paper, we describe how we solve
this, and produce a general-purpose workalike of EPICS based on Python.

\section{The submit/notify pattern for GUI programming}\label{sec:gui}

As has been mentioned above, succinctness and clarity are pursued in many
fields of programming at HEPS; GUI programming is no exception from this.
In this section, we will discuss a GUI programming pattern inspired by
EPICS, which in turn helped to inspire our Python-based EPICS workalike
besides being useful in itself.  In our summary, the complexity in GUI
programming has a few common sources, the first of which is the mixing of
business logic with operations on widgets that represent the logic.  This
kind of mix-up is a quite well-known type of non-modularity in software
engineering; it is generally resolved by decoupling GUI programs into
frontends and backends, \eg\ those in \prog{Mamba} \cite{liu2022b, dong2022}.
Another source of complexity is the complex interactions between widgets
in GUI frontends, which complicates the call chains between these widgets.
In most GUI frontends, there are some kind of quite complex states
implicitly shared between widgets, which in conjunction with the strong
concurrency usually associated with GUIs result in a significant difficulty in
understanding transitions of the implicit states.  This can easily result in
timing problems, most often occurring as race conditions; it is yet another
source of complexity, especially when coupled with complicated call chains.

\begin{figure}[htbp]\centering
\includegraphics[width = 0.6\textwidth]{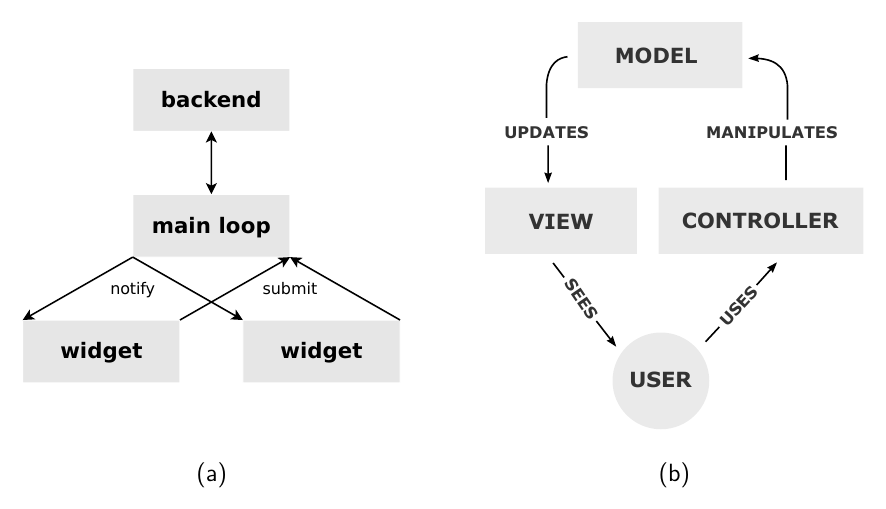}
\caption{%
	(a) The submit/notify pattern, in comparison with
	(b) the MVC pattern; the latter image is courtesy of
	the Wikipedia entry ``\emph{Model-view-controller}''.
}
\label{fig:submit-notify}
\end{figure}

GUIs in the EPICS ecosystem, known as operator interfaces (OPIs), can be
easily composed in OPI editors in a drag-and-drop fashion.  As drag and drop
are inefficient in the creation of large OPIs with many repetitions, we are
also systematically exploring \prog{PyDM} \cite{slepicka2018}, the Python-based
OPI engine for this task.  Notwithstanding this issue, we still find GUIs an
advantage of EPICS, since we do not need to worry about the problems
in the previous paragraph: OPI widgets do not directly interact or share
states with each other, eliminating problems about call chains and timing.
This is because the control logic in EPICS is gathered in input-output
controllers (IOCs), and widgets are simply graphical representations of
process variables (PVs); the latter are provided by IOCs, and manipulated
by the control logic therein.  Our attempts to systematically simplify the
control logic in IOCs, by means of \prog{caproto}-based Python IOCs, will
be covered in the following sections, and here we continue our discussion
on GUIs.  Learning from OPIs, even in a general GUI we can require widgets
to communicate only with the main event loop (\figref{submit-notify}a);
to minimise state sharing, message passing can be mandated for this
communication.  To perform message passing, libraries like \prog{ZeroMQ}
and Python's \prog{queue} can surely be used; the signal/slot mechanism
provided by some GUI libraries, \eg\ \prog{Qt} and \prog{GTK}, can
also be readily used; here we additionally note that the CA protocol
can also be regarded as a kind of message-passing mechanism.

We call the design pattern in \figref{submit-notify}(a) the \emph{submit/notify
pattern}, where widgets submit update requests (\eg\ setpoint inputs from the
user) to the main event loop, and the main loop notifies widgets of actual
updates (\eg\ readback values and other status changes).  The interactions
between a widget and the rest of the program are limited to the submissions
it sends and the notifications it receives; the similar can be said for the
main loop.  In this way, the widgets and the main loop can be considered
``actors'' well decoupled from each other, as in the \emph{actor model} of
concurrent programming, which makes the program easy to reason about.  The actor
model, as well as its cousin model, \emph{communicating sequential processes}
(CSP), are highly influential: this can be seen in languages like Erlang and
Go, as well as in libraries like \prog{MPI} and \prog{ZeroMQ}.  Historically
the model was also closely related to Smalltalk, the language well known for
pioneering in object-oriented programming (OOP).  If we also take the backend
into consideration, the submit/notify pattern may also be compared with the
model-view-controller (MVC) pattern (\figref{submit-notify}b).  The ``view''
part obviously corresponds to the widgets, and the backend is the real ``model''
part; the main loop should be a thin encapsulation of the backend, and thus
also corresponds partially to the ``model'' part.  The ``controller'' part
is the submit/notify logic separated into the widgets' event callbacks
(or slot functions in the signal/slot mechanism) and the main loop.

As is hinted above, the main event loop in a GUI frontend should only contain
logic and states essential to the graphical representation of the business
logic -- which in turn belongs to the backend; this improves the reliability
and maintainability of GUIs.  In the \prog{Mamba} framework, the backend
communicates with the frontends through a remote-procedure call (RPC) mechanism
(\figref{submit-notify}a) based on \prog{ZeroMQ}.  For tasks (\eg\ some in
beamline control) that do not need the full \prog{Mamba} infrastructure or
even barebone \prog{Bluesky} \cite{allan2019}, we also developed some standalone
GUIs with the submit/notify pattern; in these GUIs, there is still a clear
separation between frontends and backends.  \prog{Mamba} frontends developed
in the submit/notify style are, as of now, primarily used in our framework
for automated attitude tuning of beamlines \cite{li2025}.  The code for
these frontends has been released in the open-source edition of \prog{Mamba};
released in the same repository are two \prog{PyQt}-based examples for
standalone submit/notify GUIs.  One of them is a workalike of the program
in \cite{du2012}, essentially an X-ray beam-position monitor (XBPM) based
on area detectors, with a simplified RPC mechanism between the backend and
the frontend based on Python's \prog{queue}.  The other is a simplified
workalike of the \prog{ImageJ} plugin provided by the \prog{ADViewers}
module, treating \prog{areaDetector} IOCs as its backends,
reusing the CA protocol for message passing with them.

\section{The client-server model and EPICS IOCs}\label{sec:arch}

\begin{figure}[htbp]\centering
\includegraphics[width = 0.6\textwidth]{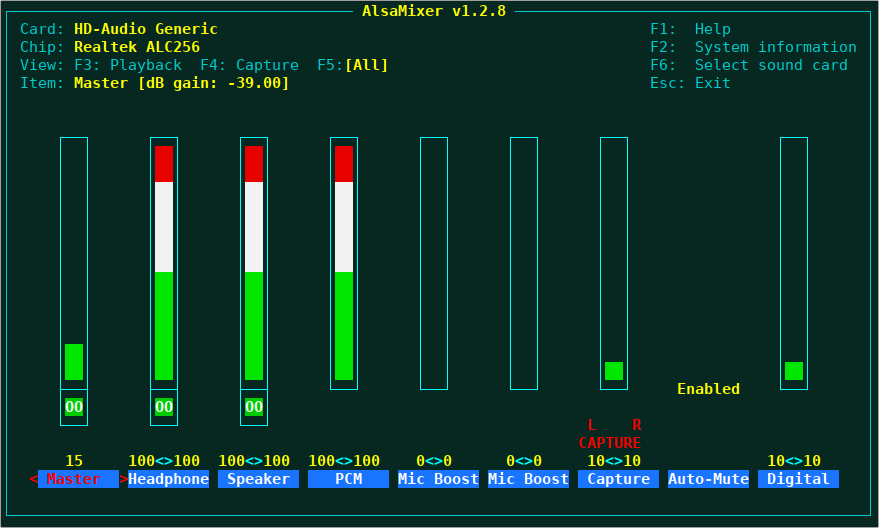}
\caption{The \texttt{alsamixer} program.}\label{fig:alsa-mixer}
\end{figure}

As has been discussed in \secref{gui}, the CA protocol can be seen as
a message-passing mechanism between servers (EPICS IOCs) and clients (EPICS
OPIs \etc); moreover, IOCs may be compared to GUI backends (whether standalone
or in \prog{Mamba}), and sometimes even be directly used as the backends.
Therefore from analysing what is common in these server-like programs, it
is possible to find better ways to do what EPICS IOCs do; historically our
standalone backends also learned from the \prog{Mamba} backend, so here
we begin with \prog{Mamba}.  Communication between the backend and frontends
in \prog{Mamba} may be categorised into \emph{requests/replies} and
\emph{notifications}, the latter necessary due to the intrinsic weaknesses
of state polling \cite{liu2022b}.  In both standalone and \prog{Mamba}
backends there are main \emph{event loops}: in \prog{Mamba} backends, event
handling (also including sending notifications) is separated into handlers
registered to the core library, which encapsulates the main event loop; in
standalone backends, there are explicit main loops that reply to requests
and send notifications.  We find the combination of an event loop, requests
(with or without replies; an example for the latter is submissions in the
submit/notify pattern, \cf\ \secref{gui}) and notifications a common pattern
in server-like programs.  In addition to GUI backends, other examples are
also abundant, \eg\ the \verb|alsamixer| program to control audio volume
under Linux (\figref{alsa-mixer}): although not written with a frontend and a
backend decoupled from each other, this program still features an event loop,
which handles mouse/keyboard events and forwards status updates from the
operating system.  We are also aware of least one proprietary device-control
product used at HEPS that explicitly uses requests/replies and notifications
over TCP/IP network for the vendor's application programming interface (API).

\begin{figure}[htbp]\centering
\includegraphics[width = 0.8\textwidth]{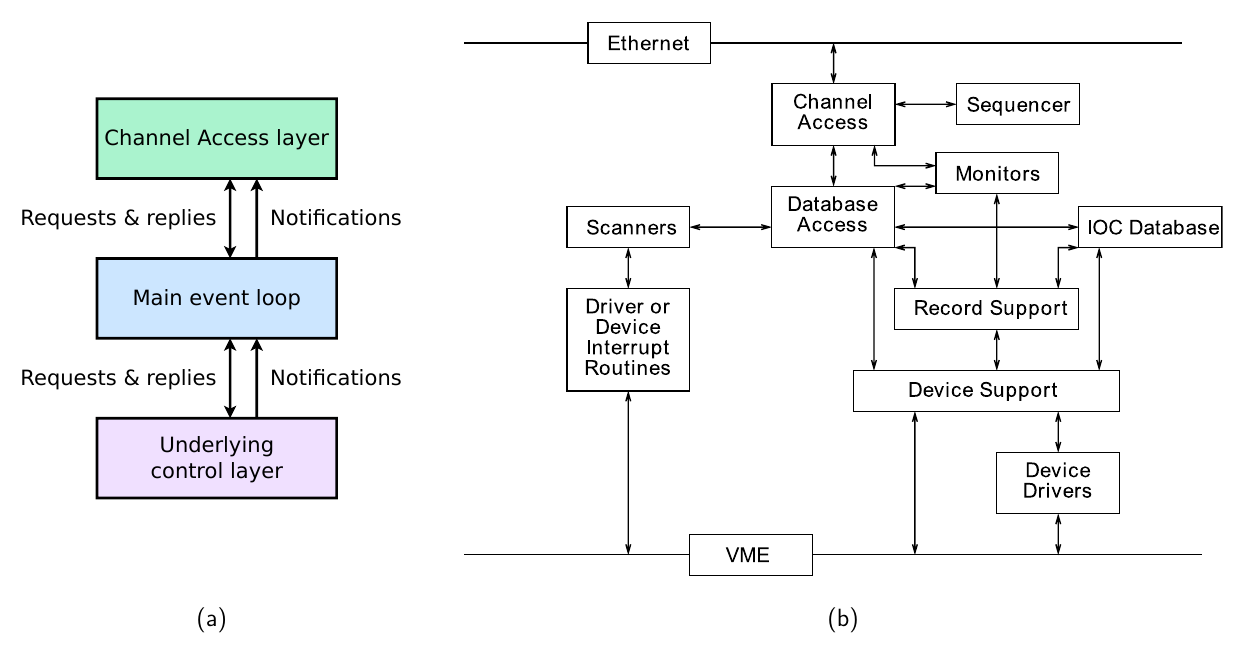}
\caption{%
	(a) The architecture of \prog{QueueIOC}, in comparison with (b)
	the architecture of an EPICS IOC; the latter image is courtesy
	of \cite{kraimer2018}.  Although (b) is for the stagnant 3.15.x
	branch of EPICS, we do not find its actively developed 7.x
	branch fundamentally different in terms of architecture (except
	for the introduction of the PVA protocol aside from CA); the 7.x
	branch is also affected by the issues summarised in this paper.%
}
\label{fig:ioc-arch}
\end{figure}

In the light of these server-like programs, it is easy to find that
\verb|caput|, \verb|caget| and \verb|camonitor|, the basic operations
in the CA protocol of EPICS, are also specialised combinations of requests/%
replies and notifications.  Given these observations, we designed the
basic architecture of our Python IOC framework, as in \figref{ioc-arch}(a).
To accommodate the more mainstream programming style both inside and outside
the Python ecosystem, the main event loop in the framework is currently based
on a regular (non-\verb|async|) function running in a regular Python thread.
Correspondingly, there is a thin layer that isolates from the main loop
the \verb|async|/\verb|await|-based code \prog{caproto} expects (\cf\ %
\secref{intro}); as the layer is based on various message queues, we
call our framework \prog{QueueIOC}.  For many IOCs in our framework
the main loops are explicit, and are quite like the main loop in the
standalone XBPM program in \secref{gui}; however, for certain types of
requirements exhibiting obvious regularity (\eg\ the \verb|QScanIOC|-based
IOCs in \secref{dev} and all IOCs in \secref{seq}), we also abstract
and encapsulate the repeated code in their main loops, so that the
developer usually only needs to care about the essentials (\cf\ Figure
\ref{fig:lower-bound}, \ref{fig:qscan-b2985} \& \ref{fig:qbumper-sim}).

Aside from \prog{QueueIOC}, (barebone) \prog{caproto} and \prog{PCASPy},
we are also aware of other attempts at using Python in the EPICS ecosystem,
\eg\ \prog{PyDevice} \cite{vodopivec2020} and \prog{pythonSoftIOC}
\cite{cobb2021}.  Instead of comparing them in detail, here we note the
most crucial difference between \prog{QueueIOC} and the rest is that it
attempts to replace most EPICS IOCs (in the narrow sense, structured like
what \verb|makeBaseApp.pl| creates) in a systematic way, while striving to
keep the amounts of code for the Python IOCs satisfactorily close to their
intuitive complexity lower-bounds.  For the latter goal, we explicitly note
that \prog{QueueIOC} focuses on fulfilling needs in succinct ways, which
are not necessarily how EPICS fulfills the same needs.  In the following
sections, concrete examples will be given for various types of applications
doable with \prog{QueueIOC}; we believe these examples can adequately show
the potential of \prog{QueueIOC}.  Here we explicitly note that what we
want is not to aggressively replace all existing EPICS IOCs, but instead
to provide a smooth transition path to more efficient alternatives that
are (protocol-wise) compatible.  We also note that motor IOCs are perhaps
a most obvious type of IOCs currently not covered by \prog{QueueIOC},
but that we do not find this a fundamental weakness.  This is because the
implementation of the \verb|motor| record contains a state machine with a
few thousands lines of code, as well as a similar amount of ancillary code,
which would take considerable human resource to port to Python even if we
omitted some less useful features.  A simple motor-like interface is indeed
provided in \prog{QueueIOC}, but this interface is, at least for now,
intended only for ``sequencers'' (\cf\ \secref{seq}) and not real motors.

To fully understand the potential of \prog{QueueIOC}, it is instructive
to compare its architecture with the architecture of EPICS IOCs (\figref%
{ioc-arch}b).  Database access, IOC database, record support and device
support are either implicit or unnecessary in \prog{QueueIOC}; device
drivers and interrupt routines are implicit in the underlying control
layer; monitors are implicit in the CA layer.  By treating other IOCs
(communicating through the CA protocol, \cf\ also \secref{seq}) as the
underlying control layer, ``sequencers'' are just a specialised kind of IOCs.
With mechanisms like \verb|QScanIOC| (\cf\ \secref{dev}) and \verb|QSlowIOC|
\cite{zhang2024}, what ``scanners'' do can also be done succinctly with
\prog{QueueIOC}.  In summary, all architectural elements in EPICS IOCs have
satisfactory counterparts in \prog{QueueIOC}; so in the long term, we
believe \prog{QueueIOC} has the potential to be eventually capable of doing
what is done with all EPICS IOCs currently used, except those IOCs that
run under special resource constraints (most prominently VME IOCs).  There
may also be performance-related concerns about \prog{QueueIOC} due to its
Python-based nature, but we find it unnecessary to worry about: with the IOC
\verb|qscan_rate| available from the supplementary materials, it can be verified
that \prog{QueueIOC} can easily offer refresh (``scan'') rates up to at least
100\,Hz and monitor rates up to at least 500\,Hz, both of which are comparable
to what EPICS IOCs are normally expected to do; with \prog{QDetectorIOC}
(\secref{dev}), \prog{QueueIOC} also offers decent detector readout performance
comparable to \prog{areaDetector}.  Our outlook can also be extended beyond
the CA protocol of EPICS: as has been noted above, \verb|caput|, \verb|caget|
and \verb|camonitor| are just specialised requests/replies and notifications;
we think similar conclusions could be made about the PV Access (PVA) protocol
added to EPICS in recent years.  In conjunction with our observations
about server-like programs, it is also a reasonable guess that similar
conclusions could be made about the communication protocols in other
device-control ecosystems, \eg\ \prog{TANGO} and \prog{Karabo}.  If these
guesses were sufficiently accurate, \prog{QueueIOC} might become a starting
point for some kind of unified device-control ecosystem, since \prog{QueueIOC}
intentionally encourages a programming style agnostic of the device-control
ecosystem: the developer usually only needs to care about requests/replies
and notifications -- not \verb|caput|, \verb|caget| or \verb|camonitor|
(\cf\ also the discussion on modularity in \secref{cases}).

\section{Device IOCs and ``soft'' IOCs with \prog{QueueIOC}}\label{sec:dev}

\begin{figure}[htbp]\centering
\includegraphics[width = 0.8\textwidth]{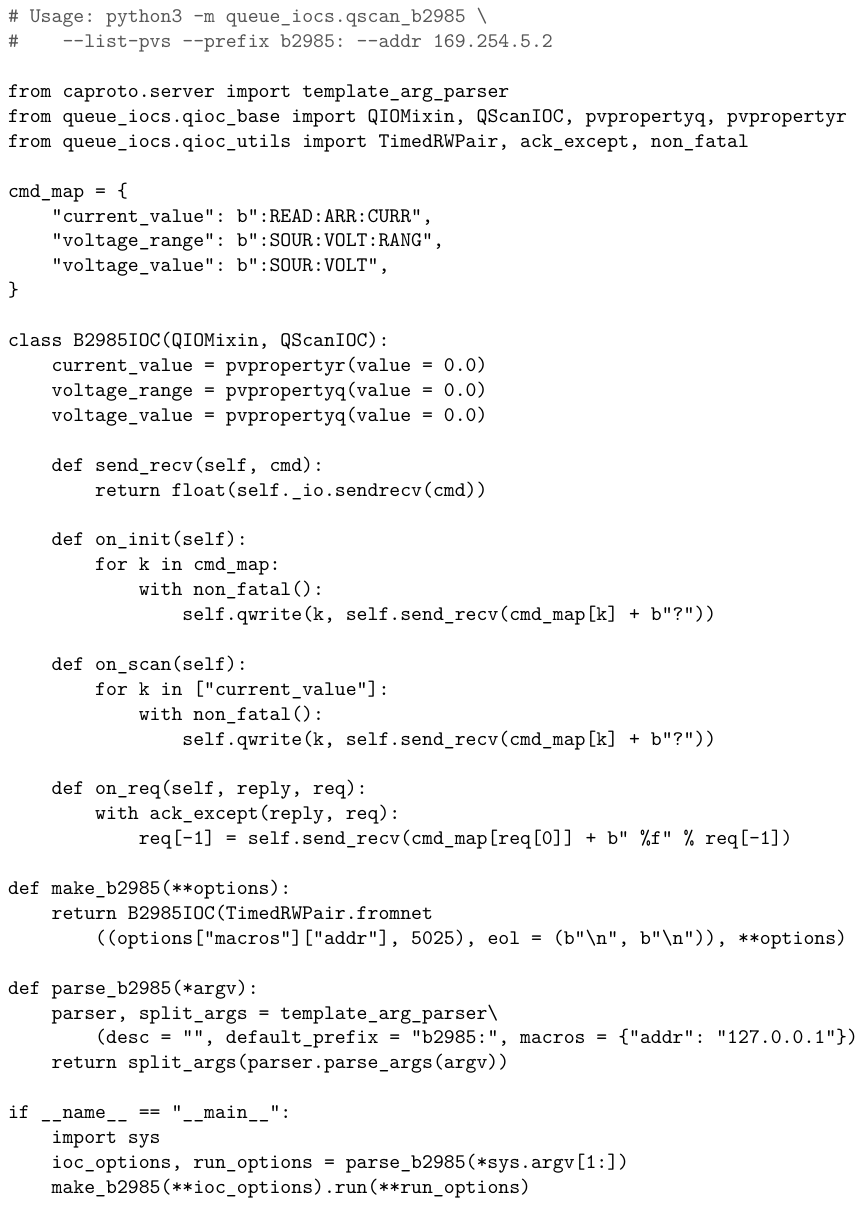}
\caption{Source code for the \texttt{qscan\string_b2985} IOC.}
\label{fig:qscan-b2985}
\end{figure}

Our actual introduction to \prog{QueueIOC} begins with a workalike of
\prog{StreamDevice}, which is in our eyes a simplest yet most useful
EPICS module: the \verb|QScanIOC| class, in combination with classes
like \verb|TimedRWPair|; an example IOC for this, \verb|qscan_b2985|, is
given for the Keysight B2985 electrometer, shown in \figref{qscan-b2985}.
As can be seen from the figure, \verb|QScanIOC| provides friendly
encapsulation for periodically polling (``scanning'') of devices,
while \verb|TimedRWPair| (just like the \prog{asyn} module) encapsulates
communication via serial ports or TCP/UDP.  Unlike \prog{StreamDevice}
which uses ``protocol files'', essentially a still quite limiting
domain-specific language (DSL), \verb|QScanIOC|-based IOCs have native
access to Python's capability to process textual and/or binary data.  As
these IOCs also have full access to Python's expressiveness, we can easily
abstract repeated code, \eg\ with \verb|cmd_map| in \verb|qscan_b2985|,
which would be clumsy to do with EPICS databases.  A more complex example
is the \verb|qscan_hfda| IOC, which will be detailed in \secref{cases}.  
Based on the extensive support for communication protocols available with
Python (whether in standard libraries or third-party libraries), with
\prog{QueueIOC} it is also easy to write IOCs based on more complex interfaces
like JSON, HTTP, modbus \etc: \eg\ the \verb|qdet_eiger| IOC for the Eiger
detector uses the vendor's protocols based on JSON/HTTP and \prog{ZeroMQ}.
\verb|qdet_eiger| is based on \prog{QDetectorIOC}, a \prog{QueueIOC}-based
framework for detector integration presented in an accompanying paper
\cite{zhang2024}, which aims to overcome some architectural limitations of
the \prog{areaDetector} framework while still offering decent performance.

\verb|QScanIOC| is intended for devices with basically stateless interfaces,
whose main event loops only differ in the logic that can be separated into
\verb|on_init()|, \verb|on_scan()| and \verb|on_req()|.  This is why the main
loop in \verb|QScanIOC| does not need to be explicitly written for each device;
for most other devices, explicit main loops are necessary.  For training
purposes in writing main loops, the \verb|qseq_bestec| IOC in \secref{cases}
is an expressive example; the \verb|qdet_nct| IOC for the Tsuji (N)CT counter
and the \verb|qdet_o974| IOC for the Ortec 974 counter are also good examples,
both based on \prog{QDetectorIOC}.  \verb|QScanIOC| is not used as we want
them to support \prog{areaDetector}-like acquisition of a fixed (and possibly
infinite) number of frames, a feature unfit for stateless implementations.
\verb|qdet_nct| can automatically detect the number of counter channels on
the specified device, and adjust its set of PVs accordingly, without relying
on code generators (\cf\ \secref{intro}).  \prog{QueueIOC}'s automatic
adaption to variable device interfaces (\cf\ also \figref{qbumper-sim},
the \verb|qseq_bestec| IOC and the self-made SDKs in \secref{cases}) is
based on Python's mechanisms for runtime creation and using of classes;
based on these mechanisms, we can also create advanced abstractions for
variable sets of detector features, similar to those in IOCs based on
\prog{ADGenICam} but with much more succinct code \cite{zhang2024}.
As can be seen from \figref{qscan-b2985}, IOCs based on \prog{QueueIOC},
whether for devices with variable interfaces or not, are easy to run for
operators: unlike \verb|st.cmd| in EPICS, with these Python IOCs device
identifiers (addresses, ports \etc) and other tuning parameters can be
given on the command line, which is well separated from the IOCs' source
code.  Additionally, when necessary it is also quite easy to customise
IOC behaviours on deeper levels, without needing to copy entire IOC
source files: \eg\ the file \verb|helpers/st_b2985.py| shows how
to customise the behaviours of \verb|qscan_b2985| in two aspects.

\begin{figure}[htbp]\centering
\includegraphics[width = 0.8\textwidth]{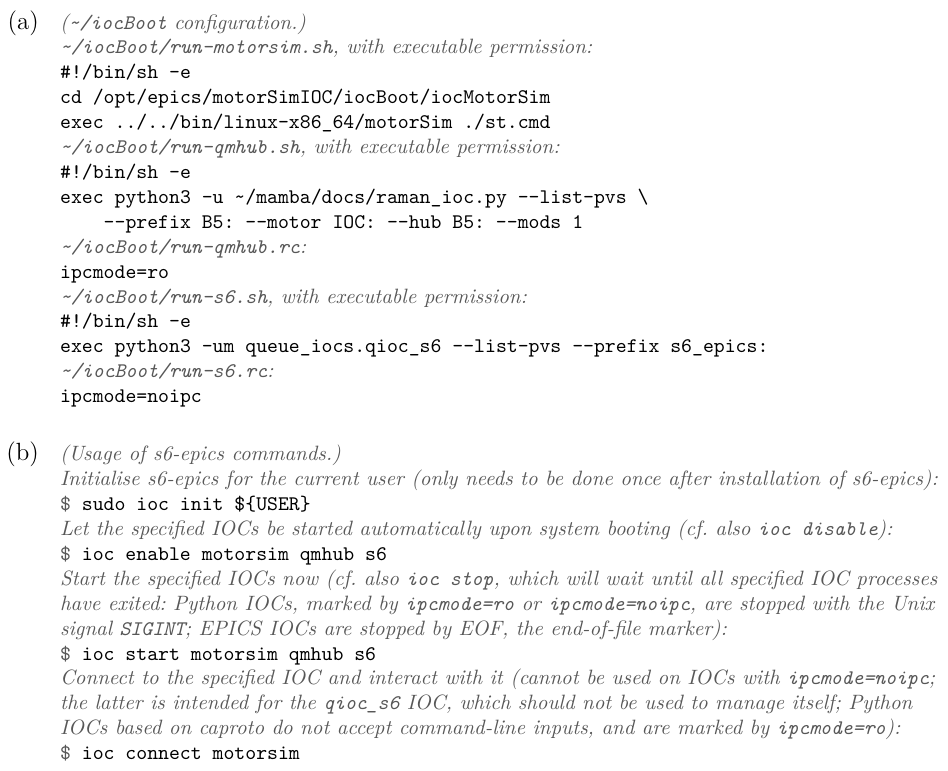}
\caption{A usage example for \prog{s6-epics}.}
\label{fig:s6-epics}
\end{figure}

Till now all IOCs discussed in this section interact with hardware other
than the controlling computer; besides ``sequencers'' which will be covered
in \secref{seq}, pure-software IOCs can also be written with \prog{QueueIOC}.
Our example for pure-software \prog{QueueIOC}-based IOCs is \verb|qioc_s6|,
a workalike for the \prog{procServControl} IOC, based on \prog{s6-epics},
a workalike of \prog{procServ} \cite{thompson2004}.  \prog{s6-epics}
itself is based on \prog{s6} (\url{https://skarnet.org/software/s6/}),
a well-designed suite of programs to manage service processes; that
latter most importantly (to us and concerning EPICS) supports separate
logging for each service with reliable log rotation to avoid exhaustion
of disk space.  After organising IOCs in the \verb|~/iocBoot| convention,
we can use administration commands like those in \cite{liu2022a}, and a few
additional useful commands (\figref{s6-epics}).  As can be seen from the
figure, aside from the ability to automatically start specified IOCs upon
booting, a currently unique feature of \prog{s6-epics} is the ability
to let both caproto-based IOCs and EPICS IOCs exit gracefully.  With
\verb|qioc_s6|, the management of IOCs controlled by \prog{s6-epics}
can be done through the CA protocol; \prog{PyDM} OPIs (\figref{atti-pydm})
have also been developed for it, available from the supplementary materials.

\section{``Sequencer'' IOCs with \prog{QueueIOC}}\label{sec:seq}

In EPICS, ``sequencers'' based on the \prog{seq} module are used to
implement modules like \prog{optics}, \prog{sscan} \etc; meanwhile, as will
be detailed in \secref{cases}, writing \prog{seq}-based ``sequencers'' can
be error-prone.  With \prog{QueueIOC} the implementation can be much cleaner,
by leveraging the expressiveness of Python and following a succinct approach.
We would also like to note that as has been hinted in \figref{ioc-arch}(b),
a less noticed aspect of ``sequencers'' is the automated manipulation of PVs
in other IOCs through the CA protocol; in our humble opinion this, instead of
state transitions, is the real essence of EPICS ``sequencers''.  Consequently,
in \prog{QueueIOC} although there is a class \verb|QSequencerIOC| for general
state machines (\cf\ also the \verb|qseq_bestec| IOC in \secref{cases}), all
``sequencers'' discussed in this section are actually based on its subclass
\verb|QMotorSeqIOC|.  The latter creates state machine with only two states,
``down'' and ``up'', in order to implement a workalike of \prog{seq}'s
``all channels connected \& received 1st monitor''.  As the name implies,
it is mainly used to implement ``sequencers'' IOCs that (just like those
in \prog{optics} and \prog{sscan}) manipulate motor IOCs, which we think
is a most common application scenario for them.  Similar to their EPICS
counterparts, all these IOCs expose some motor-like interface, which we
have intentionally designed to be unified; \prog{PyDM} OPIs (\figref%
{atti-pydm}, available from the supplementary materials) are given
for them, along with a \prog{Bluesky} encapsulation (available as
\verb|QueueMotor| in the open-source edition of \prog{Mamba}).

\begin{figure}[htbp]\centering
\includegraphics[width = 0.9\textwidth]{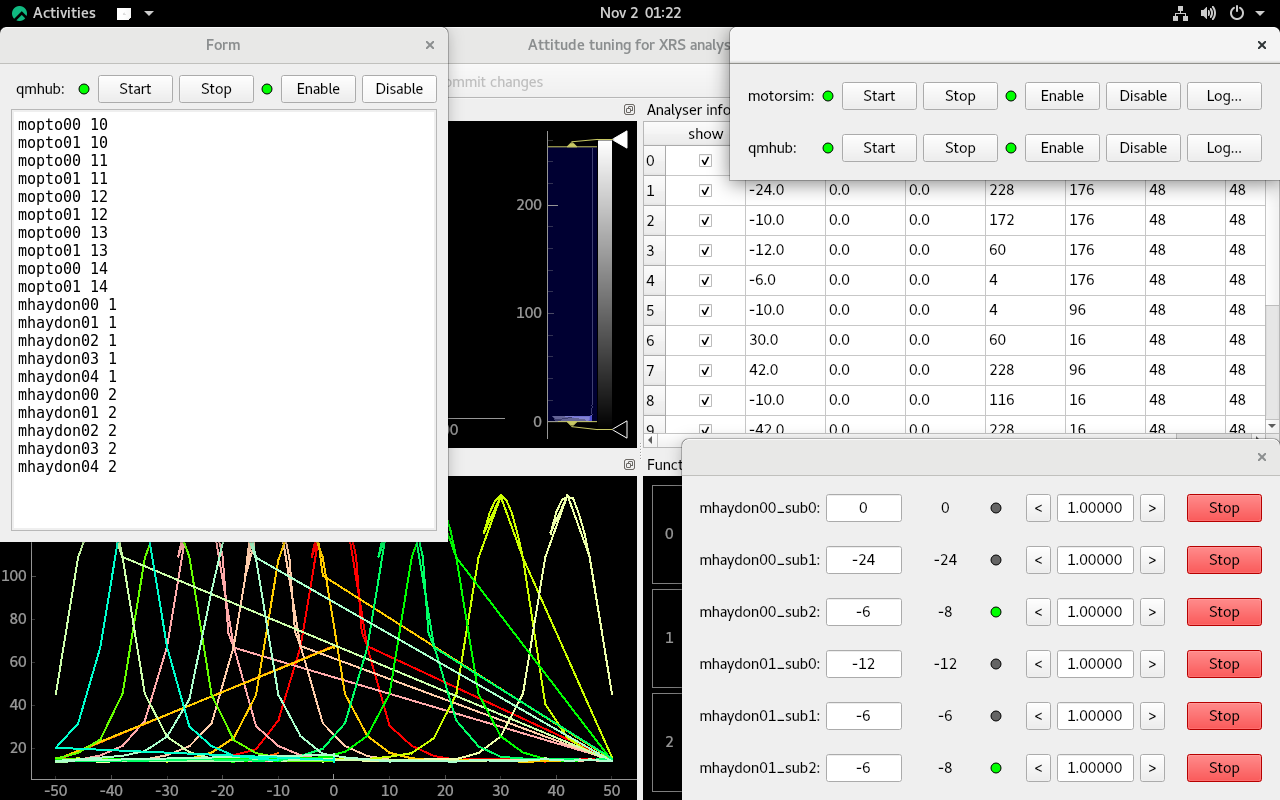}
\caption{%
	\prog{PyDM} OPIs for the \texttt{qioc\string_s6} IOC and
	\texttt{QMotorSeqIOC}-based IOCs, used in attitude tuning
	of a simulation of the Raman spectrometer at B5 of HEPS.%
}\label{fig:atti-pydm}
\end{figure}

\begin{figure}[htbp]\centering
\includegraphics[width = 0.8\textwidth]{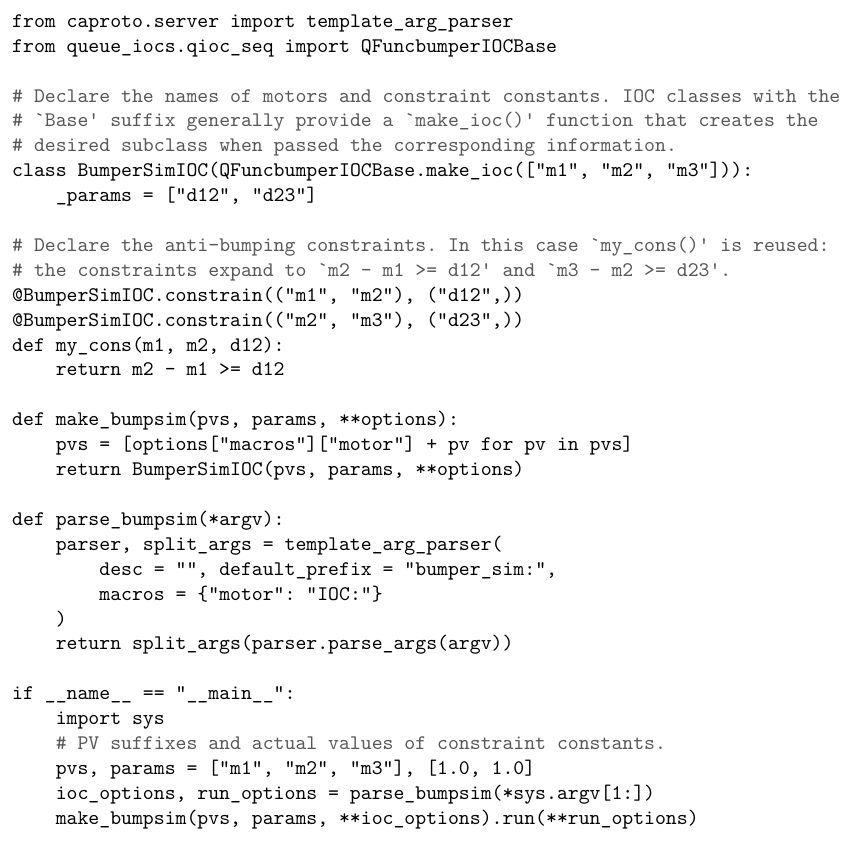}
\caption{Annotated source code for the \texttt{qbumper\string_sim} IOC.}
\label{fig:qbumper-sim}
\end{figure}

The simplest ``sequencer'' IOCs are used to systematically prevent collision
(bumping) between motors; they are based on the class \verb|QFuncbumperIOCBase|,
with \verb|qbumper_sim| (\figref{qbumper-sim}) as a simple example IOC.
A more complex \verb|QFuncbumperIOCBase|-based IOC has been deployed at the
transmission X-ray microscope beamline (BE) of HEPS, while more similar IOCs
are expected to be deployed at quite a few of the 15 beamlines of HEPS Phase I.
As can be seen from \verb|qbumper_sim|, the complicated logic in the main loop
of these ``sequencers'' IOCs are abstracted in the library, so that developers
only need to specify the essential information; in the case of anti-bumping, the
information is the mathematical constraints and the list of motor PVs.  Under
the hood, \verb|QFuncbumperIOCBase| maintains the state machine in accordance
with the underlying motor IOCs; based on this, it denies motion requests when
the specified constraints would be violated, when any motor correlated with the
requested motor (including itself) is already moving, and when the machine is in
the ``down'' state.  Depending on the requirements for other \prog{QueueIOC}-%
based ``sequencer'' IOCs, the information that needs to be specified by the
developer can be less mathematical; among them are IOCs based on the class
\verb|QMotorHubIOCBase|, used to control motors connected to multiplexer PLCs
(``motor hubs'') for motion controllers.  An example IOC based on the class
is \verb|qmhub_b5|, which is actually used in attitude tuning \cite{li2025}
of the Raman spectrometer at the hard X-ray high-resolution spectroscopy
beamline (B5) of HEPS, with a simulated variant in the open-source edition of
\prog{Mamba} (\cf\ also Figure \ref{fig:s6-epics}--\ref{fig:atti-pydm}).  A
particularly notable feature of \verb|QMotorHubIOCBase| is a delay between the
latest motion of a motor and the switch from it to any other motor connected to
the same multiplexer as this motor, along with another delay after this switch
before the newly chosen motor can be moved; both delays are tunable, and are
enforced to prevent potential power surges that may damage the devices involved.

Systematic support in \prog{QueueIOC} for monochromators, a perhaps most widely
known application scenario for ``sequencer'' IOCs, is provided in the class
\verb|QMonochromatorIOCBase|.  Based on it, workalikes of the \prog{optics}
IOCs for double-crystal monochromators and high-resolution monochromators
are respectively given as \verb|qmono_dcm| and \verb|qmono_hr|; a usage
example for the latter is given in the supplementary materials.  For training
purposes, a simple ``monochromator'' IOC is provided as \verb|qmono_sim|,
which involves two axes and simply uses their sum or difference as the
``energy'' value.  Apart from the basic features, \verb|QMonochromatorIOCBase|
also provides support for automated speed tuning to make the beam change
smoothly when the energy value is changed; of course, just like its counterpart
in the \prog{optics} IOC, this class only implements a linear approximation
to the ideal behavior.  In \verb|qmono_dcm| and \verb|qmono_hr|, support is
also provided for the user-friendly specification of Miller indices and
lattice parameters.  This feature, as well as the speed-tuning feature above,
are also provided in ways we believe to be most friendly for IOC developers:
\eg\ the latter is enabled by default, and does not need any code to be
written by the developer; it can also be opted out by simply overriding the
\verb|_auto_velo| member of \verb|QMonochromatorIOCBase|.  As can be seen
from the source code of these IOCs, in our eyes the essence of monochromators
IOCs are coordinate transformations between the energy value and the
motor positions; calibration is the process of setting up parameters for
the transformations before the latter are put into actual use.  Therefore
\verb|QMonochromatorIOCBase| can also be used for multiple-to-multiple
coordinate transformations: \eg\ those between the reciprocal and direct
spaces in crystallography, where the transformation parameters may
be imported from specialised programs like \prog{xrayutilities}
(\url{https://github.com/dkriegner/xrayutilities}) and
\prog{diffcalc} (\url{https://github.com/dls-controls/diffcalc}).
By exposing transformation parameters as PVs, like what is done in the
\verb|QMonochromatorIOCBase|-based IOCs discussed above, the importing process
can be automated.  Using the PV interface, the parameters can also be exported
to other programs: \eg\ trajectory programs used by motion controllers in
variable-speed fly scans involving monochromators or the reciprocal space,
which can achieve vastly more accurate motion behaviours than what is
possible with the speed-tuning feature of \verb|QMonochromatorIOCBase|.

\section{Case studies: some controllers and ``sequencers''}
\label{sec:cases}

\begin{figure}[htbp]\centering
\includegraphics[width = 0.81\textwidth]{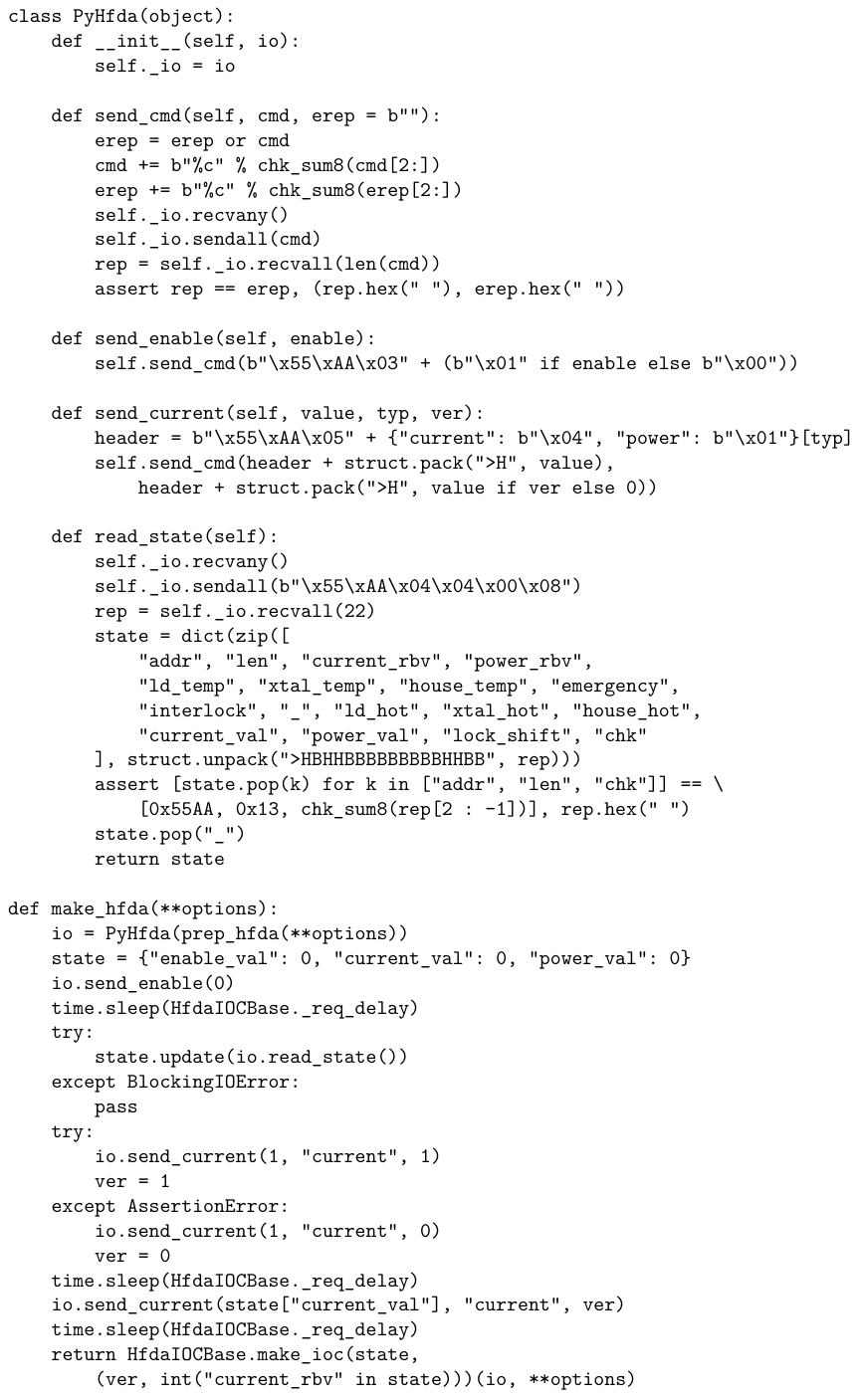}
\caption{Some notable fragments of the \texttt{qscan\string_hfda} IOC.}
\label{fig:qscan-hfda}
\end{figure}

In \secref{arch}, we noted that \prog{QueueIOC} attempts to replace most
EPICS IOCs in a systematic way, while striving to keep the new IOCs as
simple as reasonable.  We realise that after the brief tour of functionalities
in Sections \ref{sec:dev}--\ref{sec:seq}, it may still not be obvious what
the unique benefits of \prog{QueueIOC} are; so in this section, we analyse
these benefits in detail by having a close look at some example IOCs.  Our
first example is the \verb|qscan_hfda| IOC for CNI PSU-H-FDA, PSU-H-LED and
PSU-A-D laser controllers.  In this IOC, the ancillary class \verb|PyHfda| and
function \verb|make_hfda()| (\figref{qscan-hfda}) are more notable than the
IOC class \verb|HfdaIOCBase|.  While the 3 models share a basic serial-based
communication protocol, only PSU-A-D supports state readback, and there are
also minor differences in the ways to set the laser intensity; the state
readback (\cf\ the function \verb|read_state()| in \verb|PyHfda|) provides
more than 10 information items, all of which are mapped into PVs by the IOC.
The state-changing commands require checksum bytes and produce echo replies
(the models also differ slightly in their echo behaviours, as can be seen
from \verb|send_cmd()| in \verb|PyHfda|), so in case a setpoint item is not
available from the state readback, the IOC needs to decide whether to update
the value of the corresponding PV upon user writing based on the correctness
of the replies to the commands.  For this kind of readback-less setpoints
(\verb|enable_val|, as well as \verb|current_val| and \verb|power_val| for
models other than PSU-A-D), the IOC also needs to properly initialise the
corresponding hardware settings, so that the initial values of their PVs are
correctly reflected.  Moreover, the controllers may encounter read timeouts
if a state-reading command is sent too closely after a state-changing command,
so an additional delay should be enforced after the latter.  All the above
are handled succinctly and cleanly in \verb|qscan_hfda|: state readback in
\verb|read_state()| based on a call to \verb|struct.unpack()|; complete
correctness check in \verb|send_cmd()| and \verb|read_state()| based on
checksums and echo replies; the automatic differentiation between the models,
as well as the preparation of initial states, in \verb|make_hfda()|.  Done in
\verb|HfdaIOCBase| are the automatic construction of PV lists for different
models, the different handling of state changing depending on the model, and
the delay after state changing.  While these are not impossible in EPICS
IOCs, they would be very awkward to implement: what the \verb|struct.unpack()|
does above would be highly bloated with \prog{StreamDevice}; additionally
\prog{StreamDevice} cannot easily support the echo check and the delay
after state changing, and we also do not find \prog{seq}-based ``sequencers''
very helpful on this issue.  Writing the IOC in C/C++ may be also considered,
but the codebase would still be significantly larger, especially considering
the support for different versions; these make the development and
maintenance cost of such an IOC much higher than our IOC.

Admittedly, with \prog{caproto}, \prog{PCASPy} \etc, it would not be hard to
write workalikes of \verb|qscan_hfda| that are not much more complex than
it.  This is because its logic revolves around simple state-changing and
state-reading commands, and the abstraction for this pattern, \verb|QScanIOC|,
is easy to reimplement; as has been hinted above, \verb|PyHfda| and
\verb|make_hfda()| are what really make the IOC simple.  There are also
examples where the tools provided by \prog{QueueIOC} significantly contribute
to the simplicity of the IOCs: in fact, all IOCs in \secref{seq} can be seen as
examples for this; it can be easily seen when the reader considers how the same
requirements could be implemented with EPICS, \prog{caproto}, \prog{PCASPy}
\etc.  Another example is the ``sequencer'' IOC \verb|qseq_bestec| available
from the supplementary materials, used with a Bestec optical complex (PGM,
slit \& mirror unit) at the high-resolution nanoscale electronic structure
spectroscopy beamline (BC) of HEPS.  The vendor supplies a proprietary IOC-like
program that encapsulate the device interfaces as waveform PVs, which are
essentially argument lists for control commands.  As what we really want
are scalar PVs, this IOC was written to translate the argument lists
in ``master'' PVs into single elements in ``slave'' PVs, and translate
\verb|caput| operations on slave PVs back into \verb|caput| on the master
PVs.  In this IOC, a table of master-PVs and their slave-PVs is first created;
then accordingly, the slave PVs are created and the master PVs are connected
to; in the main event loop, \verb|caput| operation on the slave PVs and
monitored updates of the master PVs are handled.  This IOC is simple
and easily adaptable to another set of master and slave PVs; this is
because all the \verb|caput| and monitor events are represented in a
well-designed way there, allowing for a single function to handle all
the PVs without repeated code.  The similar requirements would be
much less convenient to implement, whether with \prog{caproto},
\prog{PCASPy} \etc\ or (especially) with EPICS and \prog{seq}.

In our summary, what distinguishes \prog{QueueIOC} from the others is an
intentional and systematic \emph{pursuit of the utmost simplicity} \cite%
{hoare1981}, or in other words approaching the complexity/cost lower-bounds
(\cf\ \secref{intro}); what \prog{QueueIOC} attempts to do, in practice, is
to provide succinct yet powerful tools that developers can use to build IOCs
that are close to their complexity/cost lower-bounds.  Here we also discuss
monochromators in detail, as another practical demonstration of how simplicity
is achieved in \prog{QueueIOC}-based IOCs -- with not only particular tools,
but also reusable design patterns.  Writing \prog{seq}-based ``sequencers''
is often error-prone, because the state transitions in them can easily become
comparable to the ``\verb|goto| hell'' C/C++ (\figref{mono-hr}c), result in
hard-to-trace bugs (a few of them are given in the supplementary materials).
This is why most ``sequencers'' in \prog{QueueIOC} only have the ``down''
and ``up'' states (\cf\ \secref{seq}); the domain-specific business logic
is instead encapsulated in the state-transition functions.  In monochromator
IOCs based on \verb|QMonochromatorIOCBase|, the functions \verb|serve_down()|,
\verb|serve_up()|, \verb|do_common()| \etc\ are the basis of our abstraction
of their business logic as coordinate transformations.  However, in addition
to these shared logic, real-world monochromators need invertible three-way
conversions between the Bragg angle $\theta$, the wavelength $\lambda$
and the energy $E$; the conversions can be even more complex for certain
monochromators, \eg\ the high-resolution monochromator (\figref{mono-hr}a).
To do this cleanly, we use functions to encode these conversions: \eg\ %
\verb|mono_phis()| in the \verb|qmono_hr| IOC (\figref{mono-hr}b),
as well as \verb|mono_theta()| which is used by \verb|qmono_dcm| and
usable by other simple variants of the double-crystal monochromator.

\begin{figure}[htbp]\centering
\includegraphics[width = \textwidth]{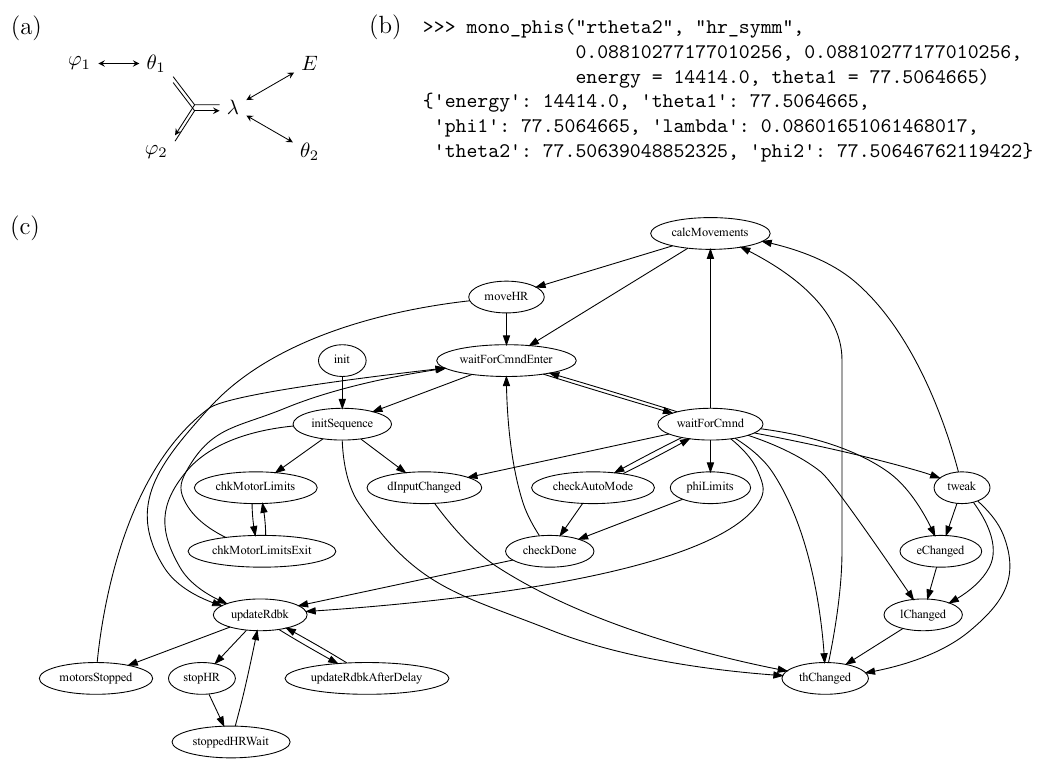}
\caption{%
	(a) Conversion between the wavelength $\lambda$, the energy $E$, the
	Bragg angles $\theta_1$, $\theta_2$ and the motorised angles $\varphi_1$,
	$\varphi_2$ of the high-resolution monochromator in its ``rock $\theta_2$''
	mode.  (b) Example usage of \texttt{mono\string_phis()}, where the 2nd
	argument specifies a symmetric monochromator geometry, while the 3rd and
	4th arguments respectively specify the Bragg spacings $2d_1$ and $2d_2$.
	(c) State transitions of the \texttt{hrCtl} ``sequencer'' in \prog{optics}.%
}
\label{fig:mono-hr}
\end{figure}

In comparison with the monochromator IOCs in \prog{optics},
IOCs like \verb|qmono_dcm| are not only much shorter, but also much
better \emph{modularised}.  The shared logic is contained in the code of
\verb|QMonochromatorIOCBase|, organised cleanly unlike the ``\verb|goto|
hell''.  The $\theta$/$\lambda$/$E$ conversions are fully encapsulated in
functions like \verb|mono_phis()|, which are implemented as straightforward
encodings of graphs like \figref{mono-hr}(a).  The rest cannot be easily
abstracted as libraries, but they are still organised cleanly, with a structure
intentionally kept consistent across different monochromator IOCs; this can
be seen from a comparison between the IOC classes, \eg\ \verb|MonoHrIOC| in
\verb|qmono_hr| and \verb|MonoDcmIOC| in \verb|qmono_dcm|.  The result of the
structural differences above is dramatic reduction in the cost of development
and maintenance.  When writing a new monochromator IOC, the developer often
only needs to customise the conversion function and the IOC class, without
worrying about the possibility of a small change conflicting with some
assumptions implicit in the global state.  When debugging, issues can usually
be easily traced and resolved thanks to the careful decoupling: \eg\ the
conversion function can isolated and used as a calculator for the variables
involved, greatly facilitating tests.  The same approach is also followed
elsewhere in \prog{QueueIOC} for non-trivial IOCs: aside from \verb|PyHfda|
in \verb|qscan_hfda|, another example is the \verb|qscan_redpower| IOC for
SPI redPOWER laser controllers based on \verb|helpers/qdev_redpower_lib.py|,
a self-made mini-``SDK''.  In the complex feature set provided by the
controllers, this SDK covers a non-trivial subset currently demanded by
the high-pressure beamline (B6) of HEPS, where those controllers are used.
Similarly, in \prog{QDetectorIOC} \cite{zhang2024}, vendors' SDKs are
also encapsulated thinly into mini-SDKs; many of these SDKs can dynamically
adapt to variable device interfaces.  These self-made SDKs expose succinct,
reusable and adaptable interfaces, which can be tested in isolation and even
reused in standalone applications.  In summary, modularity facilitates reuse,
which enhances regularity and therefore helps to reduce the complexity of the
entire system (\cf\ also the discussion on PVA, \emph{TANGO} and \emph{Karabo}
in \secref{arch}).  With measures taken like those above, the IOCs provided by
\prog{QueueIOC} have been made satisfactorily close to their complexity/cost
lower-bounds.  As has been discussed in \secref{intro}, with the consistent
pursuit of simplicity throughout \prog{QueueIOC}, we believe it is able to
bring about significant efficiency boosts in the EPICS ecosystem, and help
large scientific facilities (along with their staffs) become more intelligent.

\section{Conclusion}

Architectural deficiencies in EPICS lead to inefficiency in development and
application; from the perspective of complexity and succinctness, the essence
of these problems is that the architecture of EPICS makes the complexities of
EPICS IOCs vastly higher than the lower bounds.  A backward-compatible way to
avoid these problems is replacing EPICS IOCs with Python IOCs based on libraries
like \prog{caproto}.  Learning from EPICS OPIs, we can require widgets in GUI
frontends to communicate only with the main event loop, and mandate the use of
message passing for this communication; based on this idea, the submit/notify
pattern is formed, which is also related to the actor/CSP models and the MVC
pattern.  \prog{Mamba} frontends and standalone GUI frontends following the
pattern have been developed; the communication between both kinds of frontends
and their backends may be categorised into requests/replies and notifications,
which are handled in main event loops inside the backends.  The combination of
an event loop, requests and notifications can also be observed elsewhere; thus
by treating \verb|caput|, \verb|caget| and \verb|camonitor| as specialised
requests/replies and notifications, and handling them in Python-based main
loops, the \prog{QueueIOC} framework is formed.  After comparing the
architecture of \prog{QueueIOC} with that of EPICS, we believe \prog{QueueIOC}
has the potential to eventually replace most EPICS IOCs currently used
with succinct workalikes; under certain conditions, it may even become a
starting point for a unified device-control ecosystem.  Examples given for
\prog{QueueIOC} include workalikes of \prog{StreamDevice}/\prog{asyn}, as well
as ``sequencer'' IOCs (like those based on \prog{seq}), including those for
monochromators, motor anti-bumping and motor multiplexing; also reported are
a workalike of \prog{procServ}, as well as a \prog{procServControl} workalike
based on it and \prog{QueueIOC}.  A \prog{QueueIOC}-based framework for detector
integration, which aims to overcome some architectural limitations of \prog%
{areaDetector} while still offering decent performance, is presented in \cite%
{zhang2024}.  A practical analysis is given for the unique benefits of \prog%
{QueueIOC}, emphasising its pursuit of the utmost simplicity, which leads
to significant reduction in the cost of development and maintenance.

\section*{Statements and declarations}

\paragraph{Acknowledgements:}
Liu would like to thank the Linux Club of Peking University for providing
a venue in 2019 to share his preliminary thoughts on replacing EPICS
(\url{https://lcpu.club/wiki/index.php?title=2019\%E6\%B4\%BB\%E5\%8A\%A8B07}).
Much more importantly, LCPU introduced him to the wonderland of Unix
and Lisp, where the inspirations for this paper came from; finding
solutions much simpler than what have been previously imagined,
just like those in \cite{sussman1998} and \cite{landley2017},
has become a continuous source of joy and motivation.

\paragraph{Funding:}
This work was supported by the National Key Research and
Development Program for Young Scientists (Grant No.\ 2023YFA1609900)
and the Young Scientists Fund of the National Natural Science
Foundation of China (Grants Nos.\ 12205328, 12305371).

\paragraph{Data availability:}
\prog{QueueIOC} and \prog{s6-epics} have been released,
respectively, at \href{https://codeberg.org/CasperVector/queue_iocs}%
{\texttt{https://\linebreak[4]codeberg.org/CasperVector/queue\string_iocs}}
and \url{https://codeberg.org/CasperVector/s6-epics}; fully open-source
editions of \prog{Mamba} and \prog{ihep-pkg} have been released,
respectively, at \url{https://codeberg.org/CasperVector/mamba-ose}
and \href{https://codeberg.org/CasperVector/ihep-pkg-ose}%
{\texttt{https://codeberg.org/CasperVector/ihep-pkg-\linebreak[4]ose}}.
\prog{QueueIOC} itself depends on some patches for \prog{caproto}; GUIs
in the open-source edition of \prog{Mamba} (including the standalone ones,
which are in the \verb|mamba_lite| subdirectory) depend on some patches for
\prog{pyqtgraph}; the supplementary materials include OPIs which depend on
some patches for \prog{PyDM}; all these currently HEPS-specific patches
are available from the open-source edition of \prog{ihep-pkg}.

\bibliography{art8}
\end{document}